\documentclass[aps,pra,twocolumn,superscriptaddress,notitlepage,showpacs,showkeys]{revtex4-1}
\usepackage{graphicx,amsmath,amsfonts,amssymb,upgreek,txfonts,color}
\usepackage[colorlinks,linkcolor=blue,citecolor=blue,urlcolor=blue,breaklinks=true]{hyperref}
\usepackage[table, dvipsnames]{xcolor}
\usepackage[utf8x]{inputenc}

\begin{document}

\title{ 
	Nonlocal realism tests and quantum state tomography in Sagnac-based type-II polarization-entanglement SPDC-source
	}

\author{Ali Motazedifard} 
\email{motazedifard.ali@gmail.com}
\address{Quantum Optics Group, Iranian Center for Quantum Technologies (ICQTs), Tehran, Iran}
\address{Quantum Communication Group, Iranian Center for Quantum Technologies (ICQTs), Tehran, Iran}
\address{Quantum Sensing and Metrology Group, Iranian Center for Quantum Technologies (ICQTs), Tehran, Iran}

\author{S. A. Madani} 
\address{Quantum Optics Group, Iranian Center for Quantum Technologies (ICQTs), Tehran, Iran}
\address{Quantum Communication Group, Iranian Center for Quantum Technologies (ICQTs), Tehran, Iran}

\author{J. J. Dashkasan} 
\address{Quantum Optics Group, Iranian Center for Quantum Technologies (ICQTs), Tehran, Iran}

\author{N. S. Vayaghan} 
\address{Quantum Optics Group, Iranian Center for Quantum Technologies (ICQTs), Tehran, Iran}
\address{Quantum Communication Group, Iranian Center for Quantum Technologies (ICQTs), Tehran, Iran}

\date{\today}
\begin{abstract}
We have experimentally created a robust, ultrabright and phase-stable polarization-entangled state close to maximally entangled Bell-state with $ \% 98 $-fidelity using the type-II spontaneous parametric down-conversion (SPDC) process in periodically-poled KTiOPO$ _4 $ (PPKTP) collinear crystal inside a Sagnac interferometer (SI). Bell inequality measurement, Freedman's test, as the different versions of CHSH inequality, and also visibility test which all can be seen as the nonlocal realism tests, imply that our created entangled state shows a strong violation from the classical physics or any hidden-variable theory. We have obtained very reliable and very strong Bell violation as $ S=2.78 \pm 0.01 $ with high brightness $ \mathcal{V}_{\rm HV}= \% (99.969 \pm 0.003) $ and $\mathcal{V}_{\rm DA}= \% (96.751 \pm 0.002) $ and very strong violation due to Freedman test as $ \delta_{\rm F} =  0.01715 \pm 0.00001 $. Furthermore, using the tomographic reconstruction of quantum states together a maximum-likelihood-technique (MLT) as the numerical optimization, we obtain the physical non-negative definite density operator which shows the nonseparability and entanglement of our prepared state. By having the maximum likelihood density operator, we calculate some important entanglement-measures and entanglement entropies. 
The Sagnac configuration provides bidirectional crystal pumping yields to high-rate entanglement source which is very applicable in quantum communication, sensing and metrology as well as quantum information protocols, and has potential to be used in quantum illumination-based LIDAR and free-space quantum key distribution (QKD).

\end{abstract}




\maketitle

\section{Introduction \label{sec1}}

Generation and characterization of polarization-entangled Bell-states are very important from fundamental and applied point of view in the different contexts of the \textit{photonic}-based \textit{quantum technologies} such as quantum metrology, realizing protocols of quantum computing, and also  quantum communication (for a review, see \cite{boyd,shySPDC,lingSPDCreview2021,quantummetrologybook}). To our knowledge, the cleanest, easiest, and most inexpensive and accessible entanglement source, notably the polarization entanglement source, can be realized via the process of spontaneous parametric down-conversion (SPDC) in a bulk nonlinear crystals (NLCs) such as BBO, PPKTP or BaBO. 
However, there are different methods to generate the polarization entangled photon pairs, such as the non-post selective two-photon resonant excitation scheme on a single semiconductor quantum dot \cite{mueller2014} and the high-harmonic generation method plus a mixed co-propagating elliptically polarized waves in an isotropic media \cite{Cohen2014}. Moreover, very recently, it has been proposed \cite{guo2021} an interesting method to generate the long-distant polarization-spatial-mode hyper-entanglement for quantum communication using the PPKTP crystal over 11 km in multicore fiber. The authors have shown that the total entanglement purification efficiency can be estimated about 3-order more than the experiment using two pairs of entangled states with SPDC. 
It should be remind that the probabilistic nature of their generation process in SPDC or four wave mixing leads to creation of zero or multiple photon pairs yielding a Poissonian distribution which limits their applications such as in complex algorithms where many qubits and gate operations are required.


SPDC is the well-known nonlinear optical process, in which a classical pump laser beam at higher frequency is incident onto an optical nonlinear material which under the so-called phase-matching (PM) conditions, i.e., energy-momentum conservation, a twin-photons beam, i.e., a time-energy entangled \cite{timeenergyEntangleemnt1992} pair of signal-idler photons at lower frequencies, is generated out of quantum vacuum \cite{shySPDC,boyd}. 
One of the advantages of nonclassical sources based on the SPDC in bulk NLCs is that they are more inexpensive, user-friendly, and work at room-temperature, which means that need no cooling, and have capability to be scalable in order to be commercialized.
During these three decades, the SPDC in NLCs has been a variety range of applications in quantum metrology, quantum communication, quantum computing, quantum information, and quantum thermodynamics (for a review, see Ref.~\cite{quantummetrologybook}). 
Among these, one can remark, for example, EPR realization \cite{epr}, the entanglement generation \cite{polarizationentanglement1995,forbes,aliBellBBO,aliDNA,jin2021PPKTPsagnac,warke2021PPLNcomb}, quantum state teleportation \cite{teleport2001shih}, 
quantum ellipsometry \cite{quantumellipsometryexperimentm,quantumellipsometry2018}, quantum illumination \cite{qilluminationPRL,qilluminationScience,kadirradar,quantumilluminationbalaji1,electricFieldCorrelations,quantumradarWavefrontLanzagorta,quantumradarhelmy,quantumradarbalaji2}, quantum spectroscopy \cite{spectroscopy3,spectroscopy4,spectroscopy5,spectroscopy6}, squeezing generation \cite{squeezing1}, 
quantum imaging \cite{imaging1,imaging2,imagingshy,imagingPadgett,ghostimagingthesis},
quantum communication \cite{crypto1,crypto2,crypto3,crypto4,crypto5,crypto6,crypto8,crypto9,zbindencrypto,villoresi1,liu2002,liu2003,sheng2020,zhou2020}, and nonlocal realism tests \cite{epr,nonlocal1,nonlocal2,nonlocal3,nonlocal4,nonlocal5}.


The most important way to characterize the quantum state of the system is quantum state tomography (QST) which leads to reconstruction of the density matrices operator of the system (DMOS). Density matrice includes all information such as probabilities and coherences. Also, the evolution of the density operator enables us to evaluate the environmental effects on the quantum state of the system, and decoherence effects. 
It has been shown theoretically and experimentally \cite{kwiatstate,kwiatstate2,kwiatsource} that using the theory of the measurement of qubits, which includes a tomographic reconstruction of the DMOS due to the linear set of 16 polarization-measurements together with the numerical optimization method, the so-called \textit{maximum-likelihood-technique} (MLT), one can obtain and reconstruct the non-negative definite density operator. 
By having the density operator, all quantum coherence properties and entanglement entropies can be measured and calculated. 
Moreover, very recently, it has been theoretically developed and experimentally demonstrated a new method, resource- and computationally efficient, for quantum state tomography in Fock basis via Wigner-function reconstruction and semidefinite programming \cite{tomography2020}. It has bee shown that obtained density operator from this method is robust against the noise of measurement and relies on no approximate state displacements, and requires all physical properties. 
Furthermore, a new novel method have been developed and demonstrated for scalable on-chip QST \cite{tomographychip2018}, which is based on expanding a multi-photon state to larger dimensionality. It leads to scale linearly with the number of qubits and provides a tomographically complete set of data with no reconfigurability.

Optical interferometry is the most popular technique in optical metrology and physical optics which has been applied in a variety range of applications \cite{handbookOptic,motazediFresnel2012,motazediFresnel2014,motazediFresnel2016,motazediICTP,white1,pedroti,twowavelength}.
Among the different configurations of interferometers, in recent decades, the \textit{Sagnac}-based sources because of their stability and high rate have been used in quantum optical experiments in a variety range of application such as quantum key distribution (QKD) \cite{qKDresch,ervinQKDmaster,crypto8} and nonclassical radiation sources \cite{sagnac2004,sagnac2006,sagnac2007,sagnacThesis2010Hamel}.
Here, it worths to remind that in a Sagnac interferometer (SI), a coherent light beam is split on two branches, and thus, the two split beams pass the common-path but in opposite directions. When the two beams return to the entrance port of SI, if their wave-functions coincide, thus, coherence condition is satisfied and they interfere.

In this paper, we are motivated by the above-mentioned research to investigate the nonlocal realism tests such as Bell's inequality test and Freedman's test as well as visibility test in an ultrabirght, robust and high-rate polarization-entangled photons source via the type-II SPDC in a bulk PPKTP collinear crystal inside the common-path SI. Sagnac configuration enables us to Bidirectional pumping the PPKTP crystal inside which yields to robust and ultrabright polarization-entanglement source with high pair photon flux. After preparing and proving the entangled Bell-state, via the quantum state tomography assisted maximum-likelihood numerical optimization technique, we obtain the physical density matrix of the system, which results in entanglement entropies and entanglement measures.


\section{experimental Results} \label{secExperiment}

\begin{figure} 
	\includegraphics[width=8.7cm]{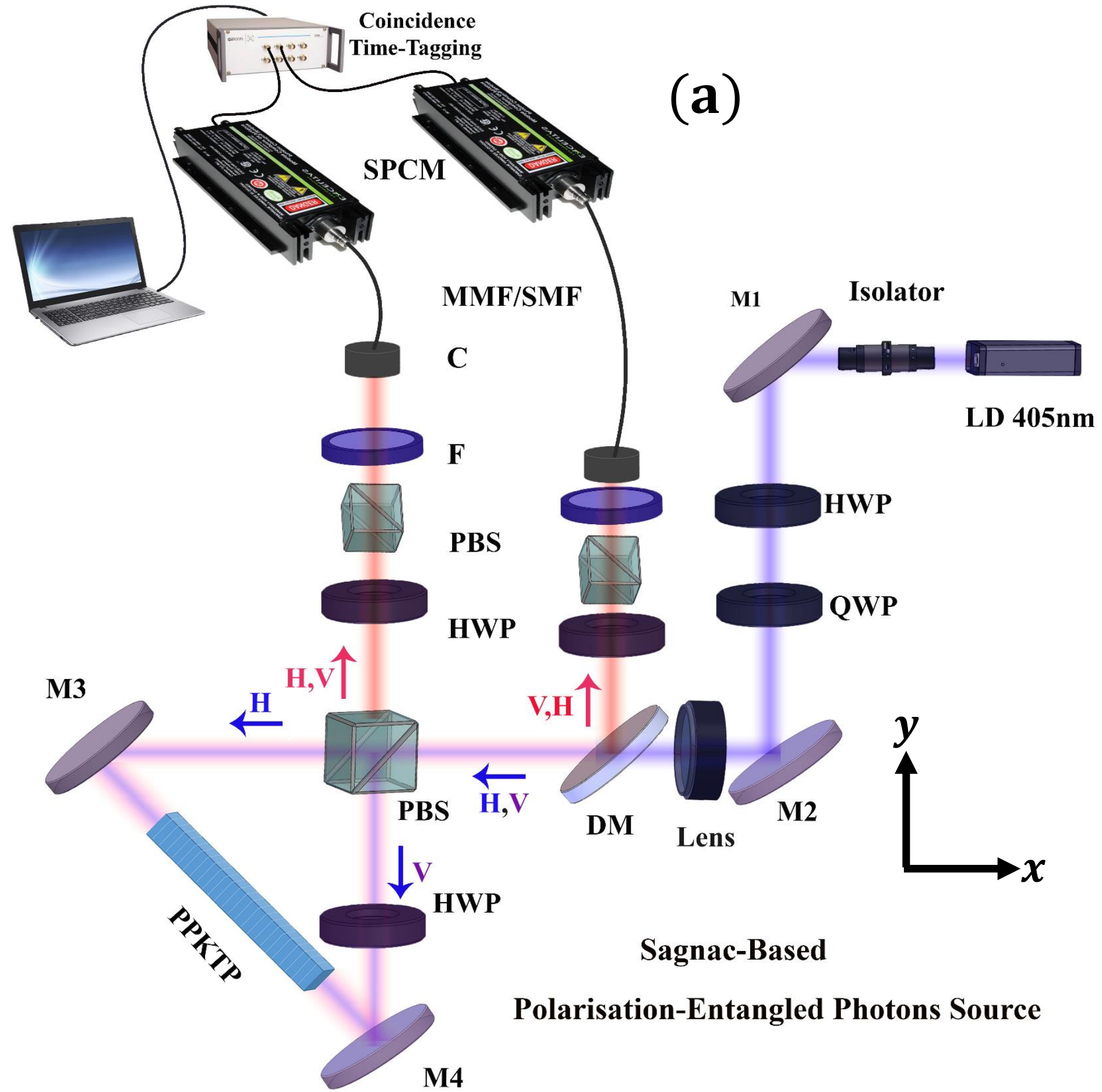}
	\includegraphics[width=8.5cm]{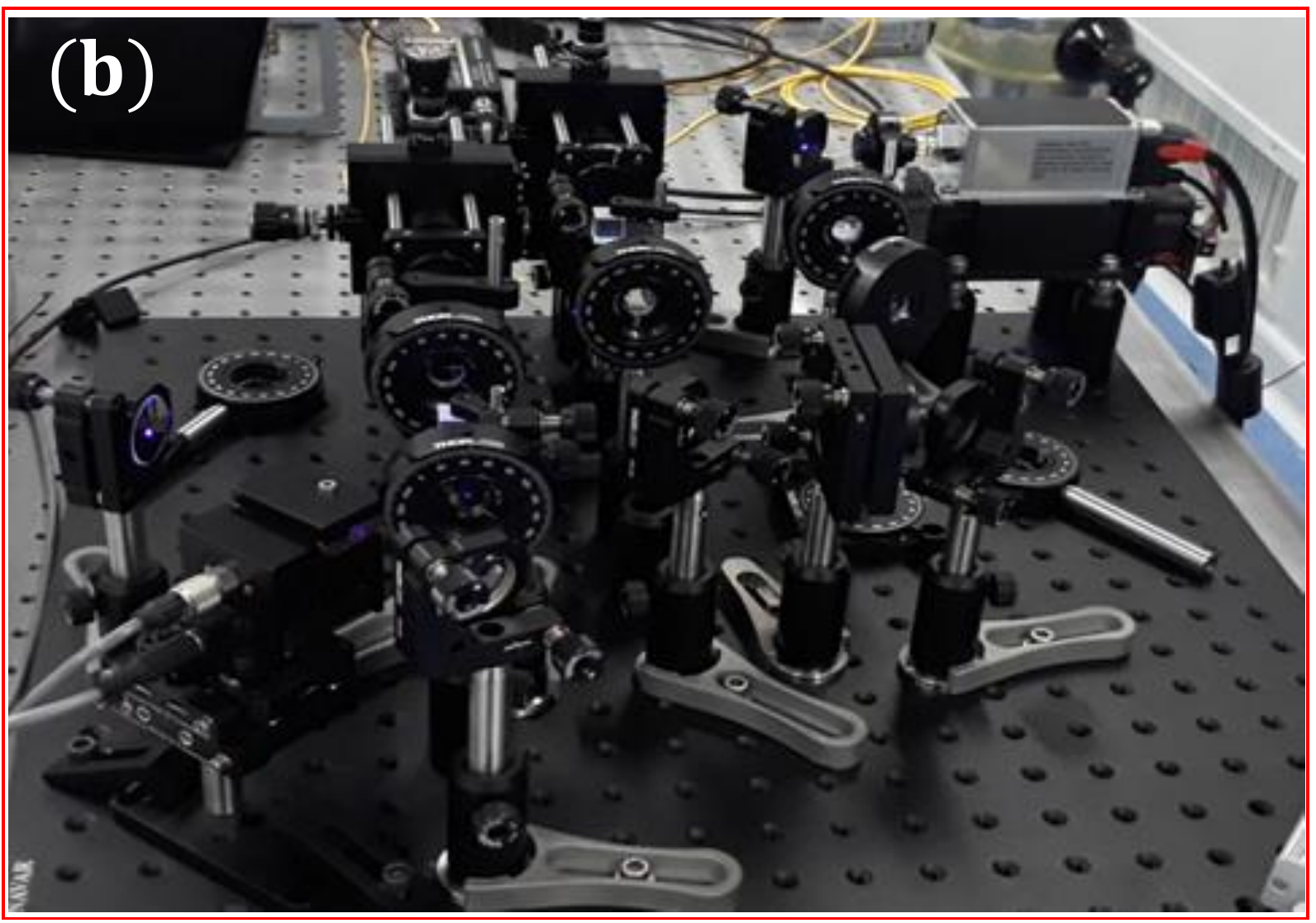}
	\caption{(Color online) (a) Schematic of SI to create the ultrabright and robust type-II polarization-entangled Bell-state at wavelength 810nm using the SPDC process in bulk PPKTP NLC. LD: 405nm diode-laser, I: optical isolator for 405nm, M1(M2): high-reflective mirror, HWP: half-wave plate [note that HWP in Sagnac arm is a dual wavelengths HWP at 405/810nm], QWP: quarter-wave plate, L: concave lens with focal length 30cm, M3(M4): high-reflective broadband mirror at 405nm and 810nm, PBS: polarizing beam-splitter [Note that in detectors arm PBS is set for 810nm while in entrance port of Sagnac a dual wavelengths PBS is used], F: 780nm-bandpass and 810nm-narrowband filters, a SPCM: Excelitas single-photon counting module, MMF/SMF moulti-/single-mode optical fiber, C: coupler lens module. Here, we used the quTools coincidence time-tagger with 81ps resolution. PPKTP: AR coted $ 1 \times 2 \times 25 ~ \rm mm $ periodically-polled-KTP collinear crystal with $ \Lambda_0=10.025\mu\rm m $ which is kept at temperature 30$ ^\circ $C by a temperature controller with 0.1$ ^\circ $C resolution. The coincidence time window and and acquisition integration time are, respectively, 5ns and 400ms. PPKTP is cut for collinear type-II SPDC QPM condition. (b) Side-view of the experimental setup.}	
	\label{fig1}
\end{figure}

\subsection{Experimental setup}

Fig.~\ref{fig1}(a) shows the schematic of experimental setup for type-II Bell-polarization-entanglement generation via the SPDC in PPKTP crystal inside the common-path SI. 
A 48mW-beam of a diode-laser passes through an optical isolator, and then, is reflected by mirror M1. In order to generate polarization entanglement in Sagnac configuration, one should bidirectionally pump the collinear PPKTP crystal with both vertical (V) and horizontal (H) polarization. 
So, to this, one should polarize the pump laser beam at $ +45^\circ $ via an HWP. Then, $ +45^\circ $-polarized laser beam is reflected by M2 toward PBS, i.e., the entrance port of the Sagnac. In the reflected/transmitted port of PBS, the V/H-polarization part of the laser beam can pass. Both V and H polarized beam are reflected by M4 and M3 in Sagnac, and then are focused on the center of PPKTP crystal by concave lens with focal length 30cm. The crystal is placed on a heater which is fiexd on a high-precision translation stage with fine-tilt capability. The crystal temperature is fixed at $ 30^\circ \rm C $ which can be controlled with $ 0.1^\circ \rm C $ precision. 
Note that to have a maximum rate and efficiency, it is necessary to focus the beam on center of the crystal which is achievable by tuning the lens and the crystal position. Each polarized beam of pump in Sagnac can generate a signal-idler pair photons at wavelength 810nm via the type-II SPDC process in the PPKTP crystal [see Fig.~\ref{fig1}(a)].

To coincide two wavefunctions $ \vert HV \rangle $ and $ \vert VH \rangle $ in order to generate the maximally polarization entangled Bell-state
\begin{eqnarray} 
&& \vert \psi^{(-)}_{\rm Bell} \rangle=\frac{1}{\sqrt{2}} (\vert H_s V_i \rangle - e^{i \phi} \vert V_s H_i \rangle ) , \label{typeIIBellState}
\end{eqnarray}
one should tune and rotate the QWP, acts as a compensator, at proper angle $ \varphi $ around the z-axis. In our experiment, to achieve the best entangled state we have set the QWP at angle $ \varphi\simeq 35.5^\circ $.
As shown in Fig.~\ref{fig1}(a), after generation of pair photons in Sagnac, pair photons $ \vert H_s V_i \rangle $ and $ \vert V_s H_i \rangle $, respectively, are reflected by the back-reflect port of PBS and dichroic mirror (DM), and then, pass through a HWP plus PBS, which act as a polarization-measurement box, and after passing through a 780nm bandpass filter and a 810nm narrowband filter with $ 12 \rm nm$ FWHM are fed via moulti-mode fibers (MMF) or single-mode fibers (SMF) to Excelitas single-photon counting modules (SPCMs) with 200cps and 400cps dark noises. The coincidence count (CC) rates are measured by quTools coincidence time-tagger with 81ps resolution. The experimental setup is shown in Fig.~\ref{fig1}(b). 
By checking the coincidence count (CC) rate at different angles of HWP in front of the SPCMs , one can tune the compensator in order to have maximum entanglement and generate an entangled state near to maximally type-II Bell-state.

\subsection{Quasi-phase matching (QPM) condition}

\begin{figure}
	\includegraphics[width=7cm]{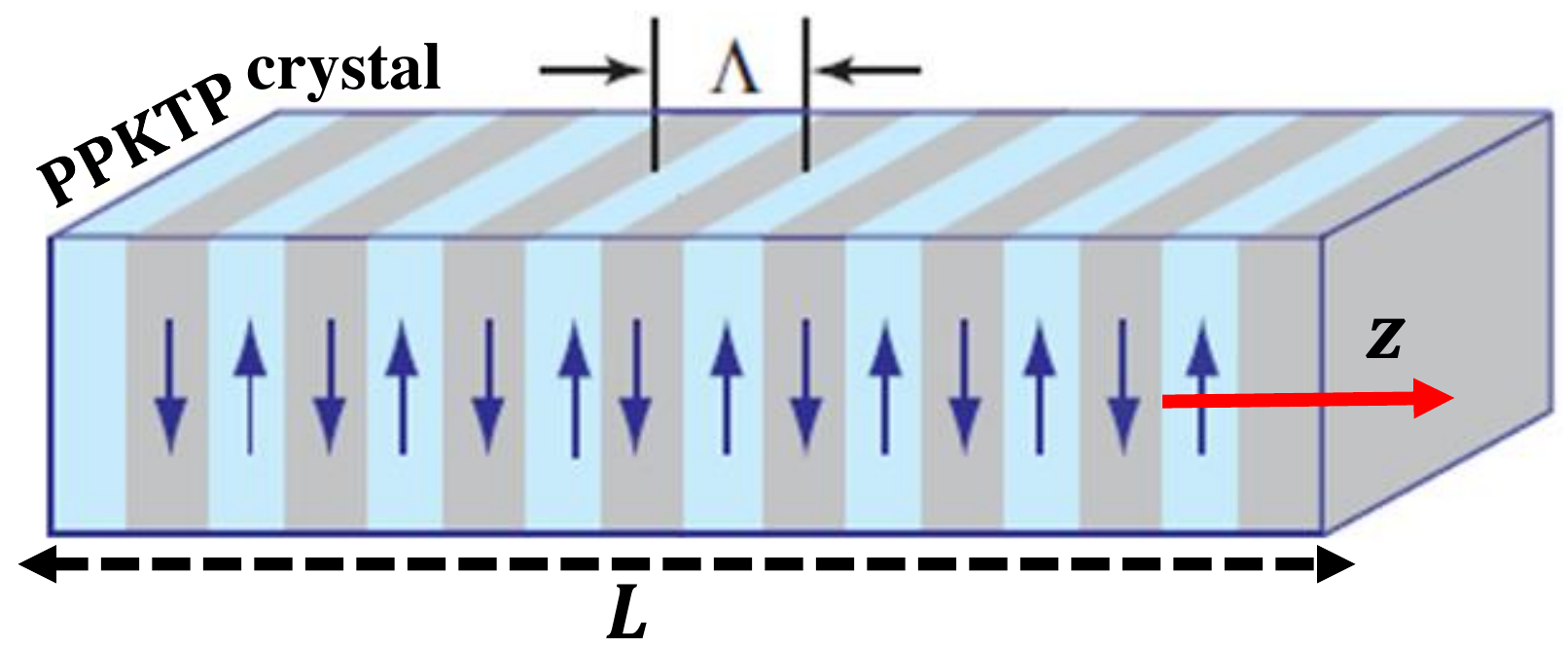}
	\caption{(Color online) schematic of a periodically-poled KTiOPO$_4 $ crystal (PPKTP) with length L. In the z-direction, the wave-vector is periodically modulated via exerting a high voltage on the coated metal on crystal with periodic length $ \Lambda $ at temperature 25$ ^\circ \rm C $.}	
	\label{fig2}
\end{figure}

Periodic polling technique \cite{periodicPolling1} is a technique based on micro-lithography for fabrication of a metallic thin film nano-layers with specified alternate wavevector orientation in a birefringent optical material in bulk or integrated optics \cite{boyd} [see Fig.~(\ref{fig2})]. 
The period $ \Lambda $ which spaces regularly the domains is multiple of the desired operation wavelength.
The periodically poled crystals, for example potassium titanyl phosphate (KTP), lithium niobate (LiNbO$ _3 $), and lithium tantalate (LiTaO$ _3 $), are frequently used as an efficient nonlinear optical materials at second-harmonic generation (SHG) and SPDC than the other types of crystals without periodic polling. Thermal pulsing, pulsed electric field and electron bombardment, or other techniques are usually used to reposition the atoms in the lattice, creating oriented domains which can be achieved either during the growth of the crystal, or subsequently.
This structure is usually designed to achieve the so-called quasi-phase-matching (QPM) condition, a generalized energy-momentum conservation, in the material to extend the wavelength regime to $ \mu \rm m $-wavelengths. 
In QPM condition, one should simultaneously satisfy relations
\begin{eqnarray}
&& \Delta \tilde{\bold k}_{\rm m} =\Delta \bold k - \frac{2 \pi \rm m}{\Lambda(\rm T)} \nonumber \\ 
&& \quad   =2 \pi  \Big( \frac{n_p(\lambda_p,\rm T)}{\lambda_p} - \frac{n_s(\lambda_s,\rm T)}{\lambda_s}-\frac{n_i(\lambda_i,\rm T)}{\lambda_i} - \frac{\rm m}{\Lambda(\rm T)}\Big) = 0 , \label{momentum1} \\
&&  \frac{1}{\lambda_p} = \frac{1}{\lambda_s} + \frac{1}{\lambda_i}, \label{energy1}
\end{eqnarray}
where $ \rm m=0, \pm 1, \pm 2, ... $, and $ \Delta \bold k= \bold k_p-(\bold k_s+\bold k_i) $, and $ n_j $ is the refractive index which depends on the wavelength and temperature, $ n_j(\lambda_j,\rm T) $ \cite{sPDCthesis}. As is seen, the last term in Eq.~(\ref{momentum1}) is the wavevector due to the periodic polling which its temperature dependency provides the long range of wavelengths for pair generation via the SPDC process. Note that here the crystal length $ L=L(\rm T) $ depends on the temperature (for more details, see Ref.~\cite{sPDCthesis}).

Let us remind that in a type-II ($ e \to e + o $) periodically poled PPKTP source, as a biphoton source, an extraordinary ($ e $) polarized pump photon is down-converted to two ordinary ($ o $) and extraordinary photons, the signal-idler biphoton, under the QPM condition \cite{boyd}. This type of NLC has potential to generate the type-II polarization-entangled Bell-state (for more details about the state of SPDC, see Refs.~\cite{shySPDC,aliDNA,aliBellBBO}).

\subsection{Nonlocality tests } 
In this subsection by verifying some well-known tests, we will show that the generated entangled state violates the classical physics prediction.
In the proceeding, we will present our experimental results of visibility test, Bell parameter measurement and Freedman test as the nonlocal realism tests for the generated state via the SPDC in PPKTP crystal in Sagnac configuration.

\subsubsection{\textbf{Bell inequality measurement and Visibility}}

We are interested in measuring CHSH (Clauser, Horne, Shimony, and Holt) inequality \cite{chshbell} which is slightly different from the original Bell's inequality. In 1964, Bell showed that all hidden-variable theories (HVTs) obey his inequality, while quantum mechanics, for example entangled state, violates \cite{bell,aspectbell,aliBellBBO}.

Bell inequality constrains the degree of polarization correlation under measurements at different polarizer angles $ \alpha, \beta $. Bell parameter $ S $ which is criterion for Bell inequality includes measure $ E(\alpha,\beta) $ which incorporates all possible polarization CC-measurement outcomes, and varies from $ -1 $ (when the polarizations always agree) to $ +1 $ (when the polarizations always disagree). It is given by \cite{chshbell,bell}
\begin{eqnarray} \label{ecorrelation2}
&& \!\!\! \!\!\!\! E(\alpha,\beta)= \frac{N(\alpha,\beta) +N(\alpha_{\perp},\beta_{\perp}) - N(\alpha,\beta_{\perp}) - N(\alpha_{\perp},\beta)}{N(\alpha,\beta) +N(\alpha_{\perp},\beta_{\perp}) + N(\alpha,\beta_{\perp}) + N(\alpha_{\perp},\beta)} , 
\end{eqnarray}
where $ o_\perp=o + 90^\circ $ (for more details and for a short review, see Refs.~(\cite{bell,aliBellBBO})).
Thus, $ S $ parameter is given by \cite{chshbell,bell}
\begin{eqnarray}
&& S= \vert E(a,b) - E(a,b')\vert + \vert  E(a',b) + E(a',b')\vert ~ ,
\end{eqnarray}
where $ a,b,a',b' $ stand for the different polarizer angles. To obtain $ S $, one needs to 16 CC-measurement. 
The most important feature of $ S- $parameter is that it is theory-independent and has no clear physical meaning. It can be proved that for any HVTs and arbitrary angles, Bell parameter satisfy inequality $ \vert S \vert \le 2 $ which is equivalent to the visibility of coincidence probability $ \mathcal{V} \le 0.71 $ in both H-V and D-A basis \cite{nonlocal4,chshbell,bell}.
Surprisingly, it can be proved that quantum mechanics can violate this inequality by considering entangled-state. Also, it can be maximized for the maximally entangled Bell-states, such state given in Eq.~(\ref{typeIIBellState}).
By choosing angles $ a=-45^\circ $, $ b=-22.5^\circ $, $ a'=0 $ (V), and $ b'=+22.5^\circ $; we obtain \cite{bell,aliBellBBO} $ S^{\rm (QM)}=2\sqrt{2}\simeq 2.81 $ for any maximally entangled Bell-state, while other quantum states give lower values such that $ S_{\! max}=S^{\rm (QM)} $. Interestingly, for these angles, it can be proved that HVT gives the maximum $ S-$value $ S^{(\rm HVT)} =2$.
The Bell inequality emphasizes that no realistic, local and complete theory, i.e., the classical physics, in the EPR context will ever agree with quantum mechanics. In the ideal case, for the Bell state (\ref{typeIIBellState}), the coincidence rate can be theoretically calculates as $ R_c^{\rm theory}= \frac{1}{2} \sin^2(\alpha-\beta)$. 
In experiment, if $ S > 2 $ ($ \mathcal{V} >0.71 $), it shows the violation the Bell inequality and approves that the nature does not agree the HVTs or classical physics.

\begin{figure}
	\includegraphics[width=7cm]{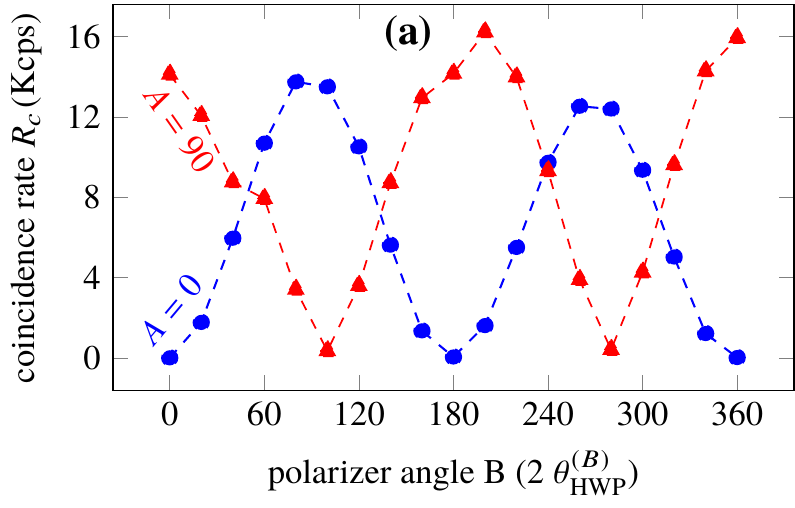}
	\includegraphics[width=7cm]{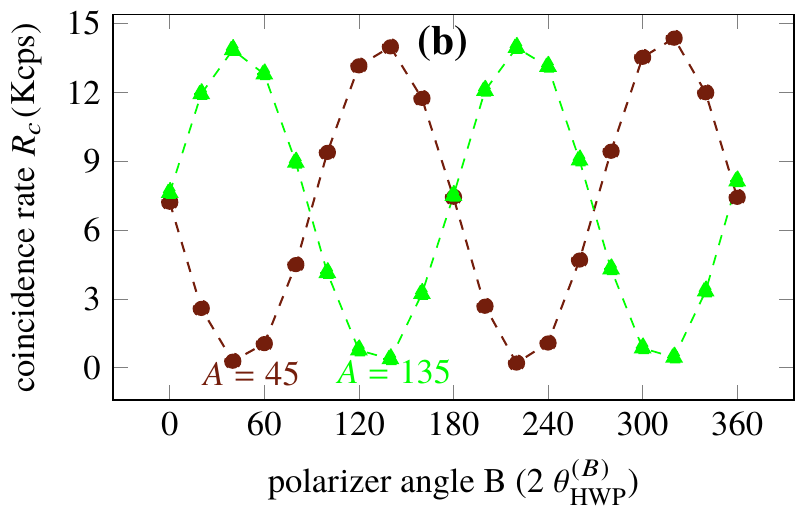}
	\caption{(Color online) Experimentally measured CC rate of channel A (1) vs. polarizer angle (the half HWP-angle) of channel B (2). (a) A is fixed at angles $ 0^\circ $ and $ 90^\circ $. (b) A is fixed at angles $ 45^\circ $ and $ 135^\circ $. Blue and brown thick-dotted points; red and green thick-triangle are referred to polarizer angles $ A=0^\circ $, $  A=45^\circ $, $ A=90^\circ $, and $ A=135^\circ $, respectively. Here, HWP plus PBS in front of detectors act as a polarizer box such that $\theta_{\rm pol}= 2 \theta_{\rm HWP}$.
	 }	
	\label{fig3}
\end{figure}

After our best tuning the state near to the maximally entangled Bell state, we start to measure the CC rate to obtain the visibility and $ S-$parameters which will be given in the following.

In Fig.~(\ref{fig3}), we have plotted the behavior of the experimentally measured CC rate when the polarizer angle of channel A(1) [or half of the HWP angle] is fixed at angles $ 0^\circ $, $90^\circ $, $ 45^\circ $ and $ 135^\circ $ for the different values of the polarizer angles of channel B(2) which agrees with the theoretical quantum prediction.
Tab.~(\ref{visibilityTable}) shows the experimental results for visibility measurement. Using the experimentally measured data, we find $ \mathcal{V}_{\rm HV}= \%(99.969 \pm 0.008) $ and $ \mathcal{V}_{\rm DA}= \%(96.751 \pm 0.002) $ with 5-digit accuracy for the visibility in H/V and 45/135 (D/A) bases, respectively. Our measured visibility shows very strong and reliable violation from the classical physics prediction ($ \mathcal{V}\le \% 71 $). Our measured visibility implies on that our polarization-entanglement source is a high-brightness nonclassical source.

\begin{table}
	\caption{ Results for experimental visibility measurement. We obtain $ \mathcal{V}_{\rm HV}= \%(99.969 \pm 0.008) $ and $ \mathcal{V}_{\rm DA}= \%(96.751 \pm 0.002) $ for the visibility in H/V and 45/135 bases, respectively.} 
	\begin{ruledtabular}
		\begin{tabular}{cccc} 
			$ \theta_{\rm pol}^{A} $ & $ \theta_{\rm pol}^{B} $ & $ R_c$ (Kcps) &  $ \Delta R_c$ (Kcps) \\ \hline \hline
			
			0 & 0 & 0.5376 & 0.0361  \\ 
			0 & 90 & 14.7923 & 0.1848  \\ 
			90 & 0 & 14.7830 & 0.1834  \\ 
			90 & 90 & 0.2919 & 0.0120 \\ \hline

			45 & 45 & 0.2743 & 0.0287  \\ 
			45 & 135 & 15.3375 & 0.1781  \\ 
			135 & 45 &  0.2211 & 0.0292 \\ 
			135 & 135 & 14.6600 & 0.2496 \\
			
		\end{tabular}
	\end{ruledtabular}
	\label{visibilityTable}
\end{table}

To calculate the Bell parameter, one should measure 16 CC rates which are given in Tab.~(\ref{tableBell}). The experimental measured rates for our prepared entangled state leads to Bell parameter $ S=2.78 \pm 0.01 $, which shows very reliable and strong violation. 
This strong violation shows a high-degree of entanglement of the prepared state in our experiment which can be used as a strong nonclassical source to quantum sensing and metrology in order to enhance the signal-to-noise ratio and the sensitivity of measurement by exploiting the quantum entanglement.

 \begin{table} 
 	\caption{ Experimental measured CC rates corresponding to the Bell inequality measurement. We have obtained $ S=2.78 \pm 0.01 $ which shows very reliable (with 78$ \sigma_{\rm std} $) and strong violation from the HVTs or any classical prediction. $ R_A $, $ R_B $ and $ R_c $ are , respectively, singles and coincidence count rates as a function of polarizer angles in channel A and B. Here, the integration time and coincidence time-window are, respectively, $ T=0.4 \rm s $ and $ \tau=5 \rm ns$. The accidental rates are calculated using $ \tau R_A R_B/T$.} 
 	\begin{ruledtabular}
 		\begin{tabular}{cccccc}
 			\multicolumn{2}{c}{angle \footnote{Here, angles are attributed to polarizer's angles that in our configuration are equivalent to $ \theta_{\rm HWP}/2 $. } [degree] } & \multicolumn{2}{c}{single count [Kcps] } & \multicolumn{2}{c}{coincidence rate [Kcps]}\\ 
 			
 			$ \theta_{\rm pol}^{A} $ & $ \theta_{\rm pol}^{B} $ & $R_A $ & $R_B $ &  $ R_c$ &  $ \Delta R_c$ \\ \hline \hline

 			0 & -22.5 & 207.6210 & 214.0125 & 1.7567 & 0.0724 \\ 
 			0 & 22.5 & 203.7475 & 207.4312 & 2.1062 & 0.0534 \\ 
 			0 & 67.5 & 200.1308 & 166.2038 & 10.8926 & 0.2150 \\ 
 			0 & 112.5 & 200.9170 & 174.8037 & 10.4809 & 0.1840  \\ \hline

 			-45 & -22.5 & 178.0473 & 213.5483 & 2.3775 & 0.0671 \\ 
 			-45 & 22.5 & 182.7763 & 207.4940 & 11.5540 & 0.1933  \\ 
 			-45 & 67.5 & 179.1138 & 166.5318 & 10.3532 & 0.1431 \\ 
 			-45 & 112.5 & 177.7340 & 174.5832 & 1.5611 & 0.0742  \\ \hline

 			45 & -22.5 & 175.5212 & 213.7574 & 11.4244 & 0.1974 \\ 
 			45 & 22.5 & 179.6321 & 207.3254 & 2.2608 & 0.0876 \\ 
 			45 & 67.5 & 177.1177 & 166.8515 & 1.9555 & 0.0765  \\ 
 			45 & 112.5 & 175.7228 & 175.2128 & 11.4158 & 0.1852  \\ \hline

 			90 & -22.5 & 160.5896 & 212.8264 & 12.3135 & 0.2163  \\ 
 			90 & 22.5 & 155.0147 & 206.7563 & 11.5542 & 0.1430 \\ 
 			90 & 67.5 & 154.5420 & 167.3865 & 1.8846 & 0.0741 \\ 
 			90 & 112.5 & 154.5125 & 174.6846 & 2.2487 & 0.0634  \\ 
 			
 		\end{tabular}
 	\end{ruledtabular}
 	\label{tableBell}
 \end{table}

\subsubsection{\textbf{Freedman's test}}

Beyond the standard CHSH Bell test which requires 16 measurements, the so-called Freedman's inequality \cite{freedman1972,freedman2018} requires only 3 coincidence measurements with very simpler mathematical calculations and experiment, more understandable from the philosophical aspects and simpler physical interpretation, notably for undergraduate students in quantum optics Laboratories.

Freedman's inequality can be quantified by Freedman parameter $ \delta_{\rm F} $ which is defined as (to see full details of derivation of Freedman's inequality, see Refs.~\cite{freedman1972,freedman2018})
\begin{eqnarray} \label{freedmanParameter}
&& \delta_{\rm F}= \Big \vert  \frac{N_c(\Phi_1)-N_c(\Phi_2)}{N_0^c} \Big \vert - \frac{1}{4}  ,
\end{eqnarray}
where $ N_c(\Phi) $ is the coincidence count for polarizer angles $ a $ and $ b $ such that $ \Phi=\vert \alpha-\beta \vert $. The optimum values for angles are $ \Phi_1=22.5^\circ $ and $ \Phi_2=67.5^\circ $ \cite{freedman1972,freedman2018}. 
$ N_0^c $ is the coincidence when both polarizers/analyzers are removed (note that in our experiment the combination of HWP and PBS in front of each detectors plays the role of polarizer). 
It can be proved that for any realistic and local theory such as classical physics, the Freedman parameter is always nonpositive, i.e., $ \delta_{\rm F} \le 0 $. If in an experiment Freedman parameters reliably becomes positive, $ \delta_{\rm F} > 0 $, thus, one says the HVTs or classical physics cannot interpret it. For example, it can be shown that an entangled state violates the Freedman's inequality and $ \delta_{\rm F} $ becomes positive which shows it cannot be explained by any HVTs or classical physics.  

\begin{figure}
	\includegraphics[width=7cm]{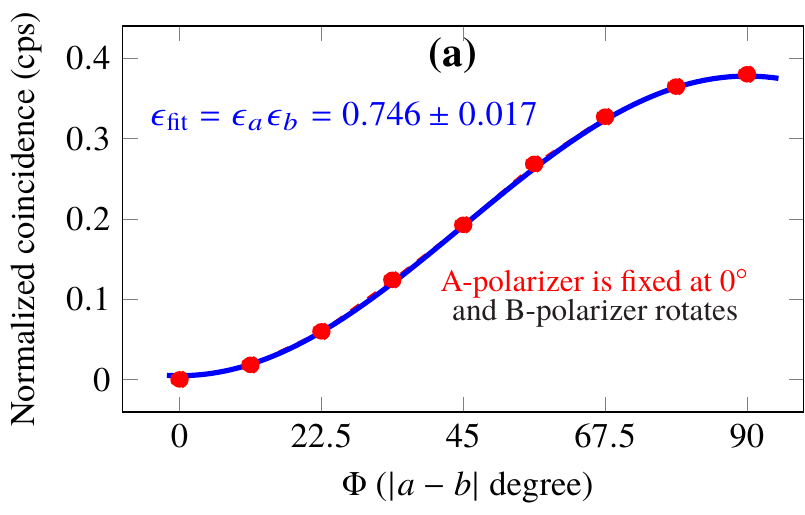}
	\includegraphics[width=7cm]{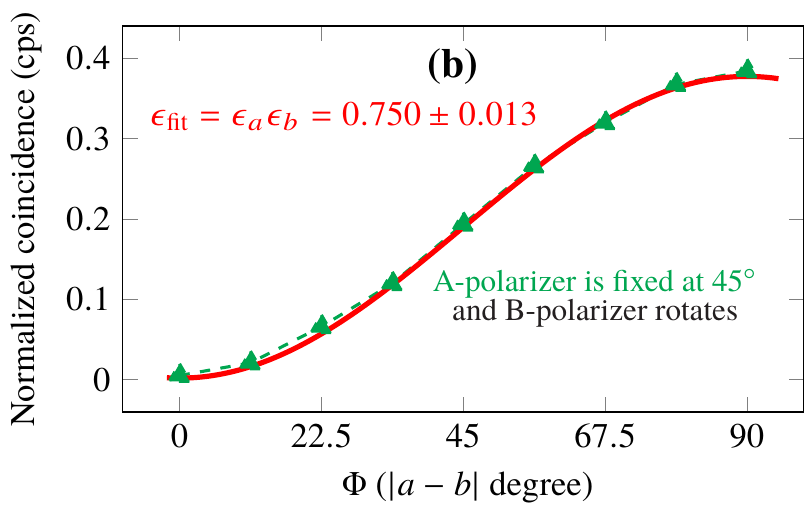}
	\caption{(Color online) Normalized coincidence count rate versus $ \Phi $ for (a) A-polarizer is fixed at $ 0^\circ $, and (b) A-polarizer is fixed at $ 45^\circ $, and B-polarizer changes such that $ \Phi $ varies zero to $ 90^\circ $. The red-dotted and green-triangle points obtained from our experiment while the blue- and red-solid lines show the theoretical fitted curve from Eq.~(\ref{n_CC_Freedman}) to the experimental points.
	}	
	\label{fig4}
\end{figure}

Let us consider Bell state $ \vert \psi^{(-)}_{\rm Bell} \rangle  $ in Eq.~(\ref{typeIIBellState}), it can be easily shown that the coincidence count can be obtained as \cite{bell,freedman2018,aliBellBBO,freedman1972,shySPDC}
\begin{eqnarray} \label{n_CC_Freedman}
&&   \frac{N_c(\Phi)}{N_0^c} = \frac{1}{2} \epsilon_a \epsilon_b \sin^2 \Phi,
\end{eqnarray}
where $ \epsilon_j $ ($ j=a,b $) is the transmittance of each arm when its polarizer box (HWP plus PBS) is present. 
Fig.~(\ref{fig4}) shows the normalized experimental measured coincidence count rate vs. difference of the polarizers angle $\phi=\vert a-b\vert $ for two cases: one angle is fixed at $ 0^\circ $($ 45^\circ $) and the other one changes. By fitting the theoretical expression (\ref{n_CC_Freedman}) to the experimental coincidence pints, we obtain the average transmittance as $ \bar \epsilon_{\rm fit}=\epsilon_a \epsilon_b= 0.748 \pm 0.015  $.

To calculate the Freedman's parameter $ \delta_{\rm F}$, we measured the CC rates for different values of $ \Phi $ [see Tab.~(\ref{tableFreedman})]. Our experimental data provide Freedman parameter with 5-digit accuracy as $ \delta_{\rm F}^{(1)}= 0.01715 \pm 0.00001 $ when $ a=0^\circ $ and $ b=22.5^\circ$ and $ 67.5^\circ $; and $ \delta_{\rm F}^{(2)}= 0.00375 \pm 0.00001$ when $ a=45^\circ $ and $ b=67.5^\circ$ and $ 112.5^\circ $. As theory predicted for entangled state for $ \Phi_{ab}=22.5^\circ $ and $ 67.5^\circ $, Freedman parameter becomes positive which shows a strong violation. It should be mentioned that to calculate the error in Freedman parameter, we have used relation \cite{freedman2018}
\begin{eqnarray} \label{freedmanError}
&& \sigma_{\delta_{\rm F}}= \Big[  \frac{N_c(\Phi_1) + N_c(\Phi_2) }{(N_0^c)^2}  +  \frac{ [N_c(\Phi_1) - N_c(\Phi_2)]^2 }{ (N_0^c)^3 } \Big].
\end{eqnarray}

\begin{table} 
	\centering
	\caption{Results for measuring the Freedman parameter $ \delta_{\rm F} $. The light-gray rows are corresponding to the maximum value of violation from the Freedman inequality. The optimized values corresponding to violation from Freedman's inequality are highlighted-gray. } 
	\begin{ruledtabular}
		\begin{tabular}{ccc|cccc}
			\multicolumn{3}{c}{ angle  [degree] } & \multicolumn{3}{c}{ rates [cps] }\\ 
			
			$ \theta_{\rm pol}^{A} $ & $ \theta_{\rm pol}^{B} $ & $ \Phi_{ab} $ & $R_A $ & $R_B $ &  $ R_c$ \\ \hline \hline \hline

			0 & 0 	&	 0	 & 202794.8 & 223713.2 & 15.50 \\ 
			0 & 11.25 & 11.25 & 203103.3 & 219832 & 741.09 \\ 
			\rowcolor{lightgray}
			0 & 22.5 & 22.5 & 202627.5 & 212547.6 & 2412.05 \\
			0 & 33.75 & 33.75 & 201121.3 & 201762.6 & 4984.87  \\ 
			0 & 45	 &	 45 & 200279 & 191063.5 & 7729.67 \\ 
			0 & 56.25 & 56.25 & 202510.3 & 179757.9 & 10767.86 \\ 
			\rowcolor{lightgray}
			0 & 67.5 & 67.5 & 202614.5 & 171022.5 & 13122.65 \\ 	
			0 & 78.75 & 78.75 & 203349.2 & 166741.3 & 14627.07 \\
			0 & 90	 & 	90 	& 204432.5 & 165988.7 & 15241.43  \\ \hline  \hline
		
			45 & 45 	& 0 	& 175596.8 & 189897.2 & 209.28 \\ 
			45 & 56.25 	& 11.25 & 173757.4 & 179289.1 & 842.59 \\ 
			\rowcolor{lightgray}
			45 & 67.5	 & 22.5 & 172162 & 171451.6 & 2645.83  \\ 
			45 & 78.75	 & 33.75 & 172518.4 & 166549.4 & 4796.4  \\ 
			45 & 90 	& 45	 & 171925.9 & 167041 & 7764.02 \\ 
			45 & 101.25 & 56.25 & 173068.9 & 170308 & 10662.76 \\
			\rowcolor{lightgray}
			45 & 112.5	 & 67.5 & 172572.9 & 178028.4 & 12819.16  \\ 
			45 & 123.75	 & 78.75 & 172656.7 & 189516.9 & 14733.98  \\
			45 & 135 	& 90 	& 174888.5 & 199757.1 & 15405.61  \\	\hline	
			
		\end{tabular}
	\end{ruledtabular}
	\label{tableFreedman}
\end{table}

\subsection{Quantum state tomography} 

\subsubsection{\textbf{density operator reconstruction}}
Now, we are going to calculate the density operator of the prepared entangled state to see how much our state is close to the maximally entangled Bell-state $ \vert \psi_{\rm Bell}^{(-)} \rangle $. 

To reconstruct the physical nonnegative density operator, we follow the generalized method of qubit measurement in Ref.~\cite{kwiatstate}. The introduced method in Ref.~\cite{kwiatstate} combines the experimentally measured different polarization-projections together a numerical optimization, the-so-called maximum-likelihood-technique (MLT), to reconstruct the physical density matrix which requires constraints (i) Hermiticity $ \hat \rho^\dag = \hat \rho $; (ii) non-negative semidefinite eigenvalue in the interval [0,1] ($ \lambda_\rho \le 1$); (iii) $ {\rm tr}(\hat \rho)=1 $; and also the important condition (iv) $0 \le {\rm tr}(\hat \rho^2) \le 1 $. In experiment, to provide the different projections, one should use a QWP before the polarizers (HWP-plus-PBS) in front of each detectors. Let us remind why only projection-measurement cannot provide the physical nonnegative density matrix. Usually in Lab due to the statistical errors and inaccuracies, experimental measured values are different from their expected theoretical ones, and thus, we cannot reconstruct the physical density operator only by experimental measuring and it needs to numerical optimization to satisfy all physical constraints. In the other words, the \textit{only} \textit{tomographic} measurement \textit{without} \textit{optimization} provides an \textit{unphysical} density matrix. 
Using the Jones matrix and our explained experimental configuration, the projection state can be defined as \cite{kwiatstate}
\begin{eqnarray} \label{projectionstate1}
&& \vert \psi_\nu \rangle \equiv \vert \psi_{\rm proj}^{(AB)}(h_1,q_1;h_2,q_2) \rangle=  \vert \psi_{\rm proj}^{(A)}(h_1,q_1) \rangle \otimes \vert \psi_{\rm proj}^{(B)}(h_2,q_2) \rangle. \nonumber  \\ 
\end{eqnarray}
It is equivalent to the projection measurement operator $ \hat \mu_{\nu}= \vert \psi_\nu \rangle  \langle \psi_\nu \vert $ with $ \nu=1,2,...,16 $. Then, the projected state of each qubit, $ \vert \psi_{\rm proj}^{(1)}(h,q) \rangle $, can be written as \cite{kwiatstate}
\begin{eqnarray} \label{projectionstate2}
&& \vert \psi_{\rm proj}^{(1)}(h,q) \rangle \! = \! U_{\rm QWP}(q) U_{\rm HWP}(h) \vert V \rangle \nonumber  \\
&& \qquad \qquad \quad  = a(h,q) \vert H \rangle +  b(h,q)  \vert V \rangle ,
\end{eqnarray}
with 
\begin{eqnarray} \label{a,b}
&& a(h,q)=\frac{1}{\sqrt{2}} [\sin 2h - i \sin (2(h-q))] , \\
&& b(h,q)=-\frac{1}{\sqrt{2}} [\cos 2h + i \cos (2(h-q))] ,
\end{eqnarray}
where $ h,q $ are, respectively, the angle of the fast-axis of HWP and QWP concerning the vertical axis-polarization.

\begin{table*} 
	\caption{ Experimental measurement data for the tomographic analysis states to reconstruct the physical density matrix using the MLT. Here, we have used the notation $ \vert D \rangle = (\vert H \rangle + \vert V \rangle)/\sqrt{2} $, $ \vert L \rangle = (\vert H \rangle + i \vert V \rangle)/\sqrt{2} $ and $ \vert R \rangle = (\vert H \rangle - \vert V \rangle)/\sqrt{2} $.} 
	\begin{ruledtabular}
		\begin{tabular}{cccc|cccc|cc}
			\multicolumn{2}{c}{ projection mode \footnote{It must be mentioned that when the measurements are taken, only one wave-plate angle has to be changed between projection measurements. That is reason to choose this set of projection $ \nu $-arrangement.} } & \multicolumn{2}{c}{ wave-plates angle} & \multicolumn{4}{c}{ single count rate [cps]} & \multicolumn{2}{c}{ coincidence count rate [cps]}\\ 
			
			$ \nu $ & $ \rm AB $ & ($ \rm h_A,q_A $) &  ($ \rm h_B,q_B $) & $R_A $  & $\Delta R_A $  & $ R_B $  &  $\Delta R_B $  & $ R_c $ & $\Delta R_A $  \\ \hline \hline

			1 & HH & (45,0) & (45,0)	& 147948.9 & 1105.93 & 153163.1 & 480.21 & 393.90 	& 15.28		\\ 
			2 & HV & (45,0) & (0,0) 	& 147634.8 & 709.19  & 166312.8 & 609.79 & 20311.20 & 154.15	\\ 
			3 & VV & (0,0) & (0,0) 		& 192185.9 & 1093.16 & 166866.8 & 935.68 & 431.10 	& 40.73		\\ 
			4 & VH & (0,0) & (45,0) 	& 191927.2 & 791.79  & 167524.8 & 1061.66& 20020.70 & 232.02	\\ \hline

			5 & RH & (22.5,0) & (45,0) 	& 167385.3 & 864.70  & 168598.9 & 678.13 & 11489.00 & 184.40	\\ 
			6 & RV & (22.5,0) & (0,0) 	& 167324.6 & 774.44  & 166527.4 & 797.42 & 9051.50  & 184.44	\\ 		
			7 & DV & (22.5,45) & (0,0) 	& 168523.2 & 828.53  & 167088.7 & 871.87 & 11123.50 & 204.01	\\ 	
			8 & DH & (22.5,45)& (45,0) 	& 168084.6 & 1171.37 & 168071.7 & 591.91 & 8941.10  & 193.73	\\ \hline

			9	&DR & (22.5,45) & (22.5,0)	 & 165216.2 & 633.79 & 164677.3 & 567.47 & 7956.20 & 228.51		\\ 
			10 & DD & (22.5,45) & (22.5,45)  & 170360   & 624.23 & 165050.3 & 776.69 & 716.90  & 56.21		\\ 
			11 & RD & (22.5,0) & (22.5,45)   & 168862.9 & 1573.20& 165050.3 & 677.33 & 9472.30 & 151.59		\\ 
			12 & HD & (45,0) & (22.5,45)	 & 142171.6 & 890.40 & 165763   & 1349.81& 10282.60& 137.01		\\ \hline

			13 & VD & (0,0) & (22.5,45) 	& 190450.1 & 1090.98 & 166000.8 & 1074.42 & 8346.60 & 110.84	\\ 		
			14 & VL & (0,0) & (22.5,90) 	& 192181.3 & 1090.04 & 168922.5 & 783.59  & 12727.60& 175.61	\\ 
			15 & HL & (45,0) & (22.5,90) 	& 146866.6 & 803.78  & 169126.4 & 784.87  & 7103.00 & 147.25	\\ 						
			16 & RL & (22.5,0) & (22.5,90) 	& 166698.1 & 1194.72 & 168640.5 & 1260.55 & 18817.20& 238.57	\\ 
			
		\end{tabular}
	\end{ruledtabular}
	\label{tabletomography}
\end{table*}

\begin{figure}
	\includegraphics[width=8cm]{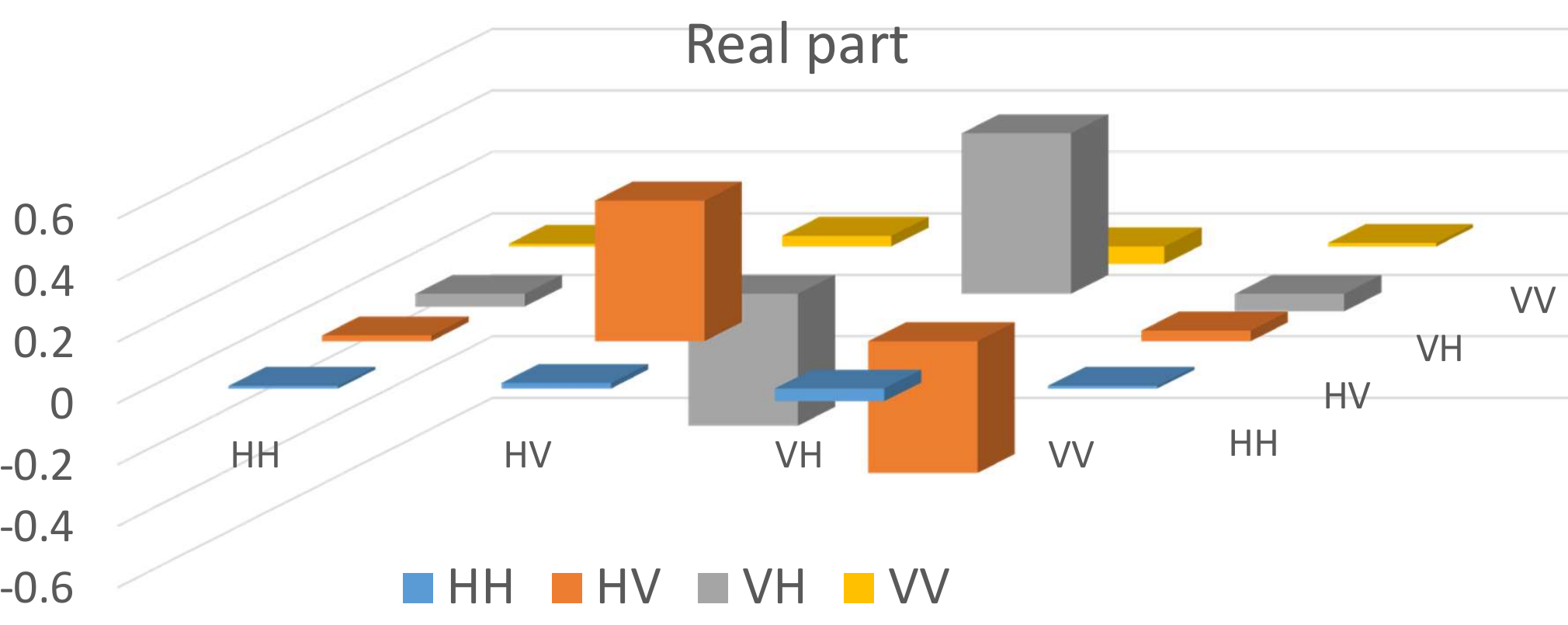}
	\includegraphics[width=8cm]{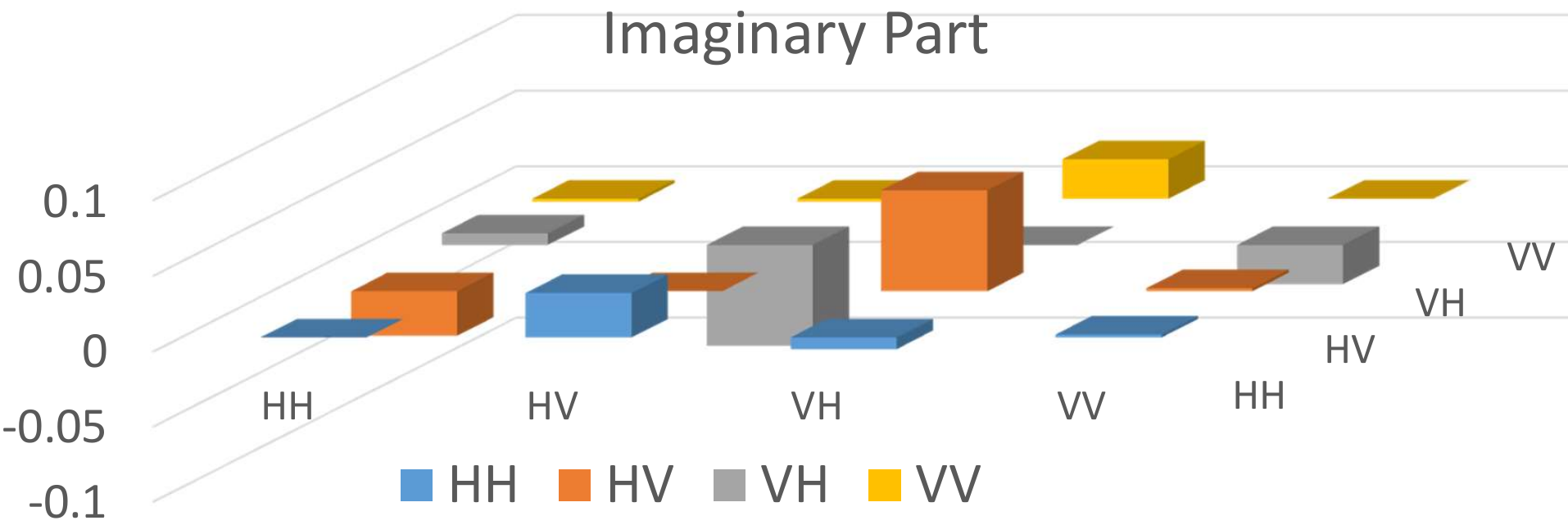}
	\caption{(Color online) Graphical representation of the real (top panel) and imaginary (bottom panel) parts of the maximum-likelihood physical density matrix obtained from the experimental tomographic data given in Tab.~(\ref{tabletomography}).
	}	
	\label{fig5}
\end{figure}

Fig.~(\ref{fig5}) shows the real and imaginary parts of the maximum-likelihood physical density matrix obtained from the experimental tomographic data in Tab.~(\ref{tabletomography}). It shows that our prepared state is close to the maximally Bell state $ \vert \psi_{\rm Bell}^{(-)} \rangle $. After numerical optimization using the measured tomographic data, the maximum-likelihood density operator matrix is obtained as
\begin{widetext}
	\begin{eqnarray} \label{roexperiment}
	\rho_{\rm rec} ^{ \rm (MLT)}= \left( \begin{matrix}
	\rm {HH} & \rm {HV} & \rm {VH} & \rm {VV}  \\
	{0.00903} & {0.0184+0.0294i } & {-0.0416-0.00769i} & {0.00875+0.00196i}   \\
	{0.0184-0.0294i} & {0.457} &  {-0.429+0.0667i} &  {0.0348+0.00201i}  \\
	{-0.0416+0.00769i} & {-0.429-0.0667i} & {0.522} & {-0.0569-0.026i} \\
	{0.00875-0.00196i}&  {0.0348-0.00201i} & {-0.0569+0.026i}  & {0.0114}
	\end{matrix} \right)_{4\times 4} \!\!\! \!\!\!\!\! , \nonumber \\
	\end{eqnarray}
\end{widetext}
with eigenvalues $ p_\rho^{(j)}= 0.93368, 0.06632 ,0,0 $ ($ j=1,2,3,4 $) corresponding to $ \rm tr \hat \rho^2 = 0.875 $.

To quantify and see how much the prepared state is close to the desired entangled state, one should calculate the fidelity $ \mathcal{F} $ in the proceeding.

\subsubsection{\textbf{Fidelity}}

Fidelity is a measure of the \textit{closeness} of two quantum states which expresses the probability that one quantum state is similar and exactly close to the other quantum state, but, it is not a metric on the density operator space. Mathematically, the fidelity for two arbitrary density operators is defined as 
\begin{eqnarray} \label{fidelity1}
&& \mathcal{F}(\hat \rho_1, \hat \rho_2)= \Big({\rm tr} (\sqrt{ \sqrt{\hat \rho_1}  \hat \rho_2  \sqrt{\hat \rho_1}})\Big)^2 .
\end{eqnarray}
It can be easily shown that the fidelity is a symmetric quantity, and thus, $ \mathcal{F}(\hat \rho_1, \hat \rho_2)=\mathcal{F}(\hat \rho_2, \hat \rho_1) $. For two pure states $ \hat \rho_j= \vert \psi_j \rangle \langle \psi_j \vert $ ($ j=1,2 $), the fidelity is simplified as $ \mathcal{F}=\vert  \langle \psi_1 \vert \psi_2 \rangle \vert^2 $, which is exactly the definition of the inner product in wavefunction space or vector space.

By considering $ \hat \rho_1=  \vert \psi_{\rm Bell}^{(-)} \rangle \langle \psi_{\rm Bell}^{(-)} \vert  $ and $  \hat \rho_2= \rho_{\rm rec} ^{ \rm (MLT)}  $, we obtain $ \mathcal{F}= \% (97.8 \pm 0.1) $ between our prepared entangled state and the maximally entangled Bell state $ \vert \psi_{\rm Bell}^{(-)} \rangle $ of Eq.~(\ref{typeIIBellState}) which shows a very good fidelity to prepare entangled state close to the desired entangled Bell state $ \vert \psi_{\rm Bell}^{(-)} \rangle $. In the following, we calculate some important entanglement entropies and measures using the reconstructed density matrix (\ref{roexperiment}).

\subsubsection{\textbf{Entanglement entropy}}

In this subsection, we are going to calculate Concurrence, the entanglement of formation, tangle, logarithmic negativity, and different entanglement entropies such as linear entropy, Von-Neumann entropy, and Renyi 2-entropy which all can be derived from the density operator.

The Von Neumann entropy which quantifies the purity of a quantum state is given by \cite{kwiatstate}
\begin{eqnarray} \label{von1}
&& \mathcal{S}=-{\rm tr} (\hat \rho {\rm log_2} \hat \rho)=- \sum_{j=1}^{4} p_a^{(j)} {\rm log_2} p_a^{(j)},
\end{eqnarray}
which for the pure state becomes zero. Our obtained Von Neumann entropy  $\mathcal{S}=0.353 \pm 0.018$, shows our prepared state is the mixture-state.
To quantify the degree of mixture of a quantum state, one can use the \textit{linear-entropy} which is given by \cite{kwiatstate}
\begin{eqnarray}
&& \mathcal{P}=\frac{4}{3} (1-{\rm tr} \hat \rho^2 )=\frac{4}{3} (1-\sum_{j=1}^{4}  p_a^{(j) 2} ), 
\end{eqnarray}
that lies between zero and 1 ($ 0 \le \mathcal{P} \le 1$). We have obtained $\mathcal{P}= 0.167 \pm 0.008$ which shows our state is not a pure state.

To compute the \textit{quantum coherence} properties of a mixed quantum state for a two-qubit system, one can calculate the concurrence, entanglement of formation, and tangle, which are equivalent measures of the entanglement of a mixed state. By considering the \textit{spin-flip} matrix $ \Sigma_f= \sigma_y \otimes \sigma_y $ as
\begin{eqnarray} \label{flip}
\Sigma_f= \left( \begin{matrix}
{0} & $ 0$ & $0$ & $-1$   \\
$ 0$& $0$ & $1 $ & $0$ \\
$0$ & $ 1 $ & $0$ & $0 $ \\
$-1$ & $ 0 $ & $0$  & $ 0 $
\end{matrix} \right) , 
\end{eqnarray}
we can define the \textit{non-Hermitian} matrix $ \hat R= \sqrt{\sqrt{\hat \rho} (\Sigma_f \hat \rho^\ast \Sigma_f) \sqrt{\hat \rho}} $ with the left and right eigenstates, $ \vert \xi_{L(R)}^{(a)} \rangle  $ and eigenvalues $ r_a^{(j)}=0.93, 0.054,0 ,0 $ with assumption $ r_1 \ge r_2 \ge r_3 \ge r_4 $, and thus, the concurrence is defined as \cite{kwiatstate}
\begin{eqnarray}
&& \mathcal{C}= {\rm Max} \left\lbrace 0 , \sum_{j=1}^{4} r_a^{(j)} {\rm sgn}(\frac{3}{2}-r_a^{(j)}) \right\rbrace , 
\end{eqnarray}
where $ \rm sgn (x>0)=1 $  and $ \rm sgn (x<0)=-1 $. We obtain the concurrence as $ \mathcal{C}= 0.876 \pm 0.007 $. 
The \textit{tangle} and the \textit{entanglement of formation } are, respectively, defined as
\begin{eqnarray}
&& \mathcal{T}= \mathcal{C}^2, \qquad  \mathcal{E}= h (\frac{1+ \sqrt{1-\mathcal{C}^2}}{2}), 
\end{eqnarray}
where $ h(x)=-x {\rm log_2}x - (1-x) {\rm log_2}(1-x) $ is a monotonically increasing function. We find that $\mathcal{T}=0.767 \pm 0.014 $ and $ \mathcal{E}=0.825 \pm 0.016$. 
\textit{Renyi 2-entropy}, which quantifies the purity of the subsystem of a pure state, is another measure to quantify the entanglement, and is defined for the density operator of a subsystem $ \hat \rho_A $ which belongs to a bipartite pure state as \cite{entropyreneyi}  
\begin{eqnarray}
&&  \Upsilon_A(\hat \rho)= - {\rm ln ~ tr} \hat \rho_A^2 , 
\end{eqnarray} 
where the reducible density operators are given by
\begin{eqnarray} \label{rhoAB}
\!\!\!\! \rho_A= \left( \begin{matrix}
{\rho_{11} +\rho_{33}} &  {\rho_{12} +\rho_{34}}  \\
{\rho_{12}^\ast +\rho_{34}^\ast} &  \rho_{22} +\rho_{44} 
\end{matrix} \right) , \quad   
\rho_B= \left( \begin{matrix}
{\rho_{11} +\rho_{22}} &  {\rho_{13} +\rho_{24}}  \\
{\rho_{13}^\ast +\rho_{24}^\ast} &  \rho_{33} +\rho_{44} 
\end{matrix} \right) .   \!\!\! \!\!\!\!\! \nonumber \\
\end{eqnarray}
The high value of Renyi 2-entropy shows a high degree of entanglement or low purity for the subsystem. In our case, we find $ \Upsilon_A= 0.684 \pm 0.014 $, which implies on impurity of our state. 
Finally, the \textit{Logarithmic negativity}, as upper bound to the distillable entanglement, which is obtained from the Peres-Horodecki criterion for separability \cite{horodeckiSeparabale}, is defined as 
\begin{eqnarray} \label{negativity1}
&& E_{\mathcal{N}}(\rho)= {\rm log_2} {\rm tr} \sqrt{\hat \rho^{\dag \rm T_A} \hat \rho^{ \rm T_A}} ,
\end{eqnarray}
where $ \rm T_A $ is partial transpose with respect to subsystem A. It should be noted that, the negativity is independent of the transposed party because $ \rho^{ \rm T_A}= (\rho^{ \rm T_B})^{\rm T} $. Therefore, if the density matrix $ \hat \rho  $ is separable, then, all eigenvalues of $ \rho^{ \rm T_A} $ are non-negative. Otherwise, if the eigenvalues are negative; thus, $ \hat \rho $ is entangled. In our case, we find $ E_{\mathcal{N}}(\rho)=0.898 \pm 0.008 $ which shows the high degree of entanglement of our prepared state. \\

\section{summary, conclusion and outlooks \label{summary}}

In this work, we experimentally prepared, measured and characterized high brightness, robust and stable type-II polarization-entangled state very close to the maximally entangled Bell-state with $ \% 98 $ fidelity via SPDC process in PPKTP NLC inside the SI. We have measured Bell and Freedman parameters as well as the visibility which are criteria for nonlocality test. They all showed the prepared state has high degree of entanglement. By reconstructing the physical density operator using the tomographic projection measurements together with the numerical optimization technique, we calculated some entanglement entropies which implied on the nonseparablity of the prepared state. 

As \textit{outlooks}, we are going to use this realized high-rate phase-stable entanglement source to implement the quantum LIDAR based on quantum illumination and to implement free-space QKD experiment.

\section{AUTHOR CONTRIBUTIONS}
 AMF proposed and developed the primary idea of the nonlocal tests in SI. Of course, NSV had chosen the Sagnac interferometer (SI) as the entanglement source for the free space QKD experiment using BB84 protocol which will be reported by our group in future soon. All authors have equally contributed to built the experimental setup and adjustments, notably JJD. But, all experimental measurements and adjustments regarding the nonlocal tests in this article have been performed by AMF and SAM. Also, the theoretical calculations and all numerical analysis have been done by AMF and SAM, respectively. JJD also drew the graphical first figure. All authors contributed to prepare the manuscript, but, AMF wrote it.

\section{acknowledgments}
All authors thank the ICQTs. Also, AMF and SAM thank Prof. Paul G. Kwiat and notably Christopher Karl Zeitler from university of illinois because of their useful comments in numerical part regarding the  quantum state tomography. 

\end{document}